\documentclass[showpacs,preprintnumbers,amsmath,amssymb,prb]{revtex4}

\usepackage{graphicx}
\usepackage{subfigure}
\usepackage{dcolumn}
\usepackage{bm}

\begin{document}

\title{Theory of field-induced quantum phase transition in spin dimer system Ba$_3$Cr$_2$O$_8$}

\author{O.\ I.\ Utesov$^1$}
\email{utiosov@gmail.com}
\author{A.\ V.\ Syromyatnikov$^{1,2}$}
\email{asyromyatnikov@yandex.ru}
\affiliation{$^1$Petersburg Nuclear Physics Institute NRC "Kurchatov Institute", Gatchina, St.\ Petersburg 188300, Russia}
\affiliation{$^2$Department of Physics, Saint Petersburg State University, 198504 St.\ Petersburg, Russia}

\date{\today}

\begin{abstract}

Motivated by recent experiments on Ba$_3$Cr$_2$O$_8$, we propose a theory describing low-temperature properties in magnetic field $h$ of dimer spin-$\frac12$ systems on a stacked triangular lattice with spatially anisotropic exchange interactions. Considering the interdimer interaction as a perturbation we derive in the second order the elementary excitations (triplon) spectrum and the effective interaction between triplons at the quantum critical point $h=h_c$ separating the paramagnetic phase ($h<h_c$) and a magnetically ordered one ($h_c<h<h_s$, where $h_s$ is the saturation field). Expressions are derived for $h_c(T)$ and the staggered magnetization $M_\perp(h)$ at $h\agt h_c$. We apply the theory to Ba$_3$Cr$_2$O$_8$ and determine exchange constants of the model by fitting the triplon spectrum obtained experimentally. It is demonstrated that in accordance with experimental data the system follows the 3D BEC scenario at $T<1$~K only due to a pronounced anisotropy of the spectrum near its minimum. Our expressions for $h_s$, $h_c(T)$ and $M_\perp(h)$ fit well available experimental data.

\end{abstract}


\maketitle

\section{Introduction}

Field-induced phase transitions in dimer spin systems have been intensively discussed in the last two decades. These magnetic systems consist of weakly coupled dimers with strong antiferromagnetic interaction between spins within a dimer. The singlet ground state in such compounds is separated by a gap from triplet excited states (triplons) at zero magnetic field that results in a spin-liquid behavior characterized by a finite correlation length at zero temperature. \cite{Mila,chern} External magnetic field $h=g\mu_BH$ lowers the energy of one of the triplet branches and the transition to a magnetically ordered phase takes place at $h=h_c$. Such field-induced phase transitions can effectively be described in terms of triplon Bose-Einstein condensation (BEC) if the system has $U(1)$ symmetry. Transitions to magnetically ordered phases are described in this way in many dimer materials \cite{chern} the most famous of which is probably TlCuCl$_3$. \cite{tlcucl3}

\begin{figure}
  \noindent
  \hfil
  \includegraphics[scale=0.6]{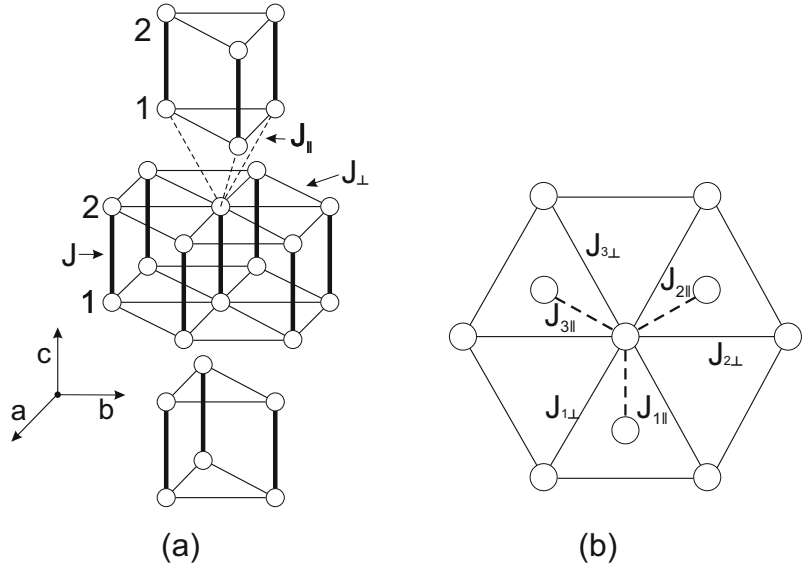}
  \hfil
  \caption{(a) Schematic crystal structure of Ba$_3$Cr$_2$O$_8$, where only magnetic atoms are presented. Solid bonds mark dimers. Exchange couplings between spins from different dimers are also shown. (b) Projection on $ab$ plane with notation of the model parameters used in the present paper.}
  \label{Ba3Cr2O8}
\end{figure}

Ba$_3$Cr$_2$O$_8$ is a dimer system of this kind which has attracted much attention recently. \cite{Kofu,KofuExp,Aczel,Dodds,aczel1,struc,zvyag,nakaj} Magnetic Cr$^{5+}$ ions have spin $S=1/2$ and form couples of triangular lattices along $c$ axis (see Fig.~\ref{Ba3Cr2O8}). \cite{aczel1,struc} Dimers are elongated along the $c$ axis. They are formed by nearest spins the distance between which is 3.93 \AA. The minimal distance between spins from different dimers is about 5.74 \AA\ and 4.6 \AA\  in $ab$ planes and between $ab$ planes, respectively, that results in a weak inter-dimer interaction. This compound has been extensively investigated experimentally by different techniques including elastic \cite{KofuExp} and inelastic \cite{Kofu,KofuExp} neutron scatterings, ESR, \cite{zvyag,KofuExp} bulk magnetization, \cite{nakaj,Aczel,KofuExp} and magnetocaloric effect. \cite{Aczel} In particular, it is found that a canted antiferromagnetic structure arises in the range of fields $h_c<h<h_s$, where $H_c\approx12.5$~T and $H_s\approx23.6$~T is the saturation field. The transition at $h=h_c$ is of the 3D BEC universality class while that at $h=h_s$ is of the first order presumably due to a spin-lattice interaction. \cite{Aczel}

Using a self-consistent Hartree-Fock-Popov approach, theoretical description is suggested in Ref.~\cite{Dodds} of the quantum phase transition to the magnetically ordered phase in Ba$_3$Cr$_2$O$_8$. Although quite good agreement is achieved in Ref.~\cite{Dodds} between the theory and the majority of experimental results, some problems remain in the theoretical description of Ba$_3$Cr$_2$O$_8$ which have to be resolved. (i) The model proposed for description of Ba$_3$Cr$_2$O$_8$ looks not realistic. \cite{Dodds,Kofu,KofuExp} It is assumed, in particular, that exchange interactions between spins from neighboring dimers are symmetric, i.e., interaction of spins 1 (see Fig.~\ref{Ba3Cr2O8}) from the neighboring dimers ('1-1' interaction for short) is the same as '2-2' interaction, and '1-2' and '2-1' interactions are also equal to each other. Whereas this assumption looks reasonable for dimers within $ab$ planes, it is most likely not the case for dimers from different $ab$ planes because the distance between spins 1 and 2 from different dimers is substantially shorter than those for '1-1' and '2-2'-couplings. (ii) The spectrum of the proposed model has a minimum at the incommensurate momentum $\approx(2.22,-2.25,2.28)$ [r.l.u.]. Thus, a helical magnetic structure would appear at $h>h_c$, whereas the neutron experiment~\cite{KofuExp} unambiguously shows the canted antiferromagnetic structure in Ba$_3$Cr$_2$O$_8$ characterized by momentum ${\bf k}_0=(\pi,\pi,\pi)$. (iii) The theory of the triplon condensation proposed in Ref.~\cite{Dodds} is semi-phenomenological because the effective interaction between condensed triplons and the stiffness of their spectrum are determined from comparison with experimental data rather than from microscopic calculations using exchange parameters.

To resolve the above mentioned problems, we propose in the present paper a more realistic model for Ba$_3$Cr$_2$O$_8$ with exchange couplings shown in Fig.~\ref{Ba3Cr2O8}. Considering the interdimer interaction as a perturbation and using the standard representation of spin operators via three Bose-operators (Sec.~\ref{trans}), we obtain spectra of triplons in this model in the second order in the interdimer interaction (Sec.~\ref{spectra}). BEC of triplons is discussed theoretically in Secs.\ref{effint} and \ref{condensation}, where expressions are derived in the second order for the effective triplon interaction at $h=h_c$, $h_c(T)$ and the staggered magnetization $M_\perp(h)$. Using our theoretical results, we describe in Sec.~\ref{appl} experimental data obtained in Ba$_3$Cr$_2$O$_8$. First, we find values of exchange constants from triplon spectrum fitting. Then, we calculate using these values $h_s$, $h_c(T)$, $M_\perp(h)$ and find a very good agreement with the corresponding experimental data. A summary and our conclusion can be found in Sec.~\ref{sum}.

\section{Theory}

\subsection{Exchange interactions and Hamiltonian transformation}
\label{trans}

It is well known that the value of the exchange interaction between spins is very sensitive to the distance between them because it is short-ranged by nature. Then, it is natural to take into account in Ba$_3$Cr$_2$O$_8$ the exchange coupling between neighboring spins only as it is shown in Fig.~\ref{Ba3Cr2O8}. As a result one leads to the following Hamiltonian:
\begin{equation}
  \mathcal{H}=\sum_i J \mathbf{S}_{i,1} \cdot \mathbf{S}_{i,2} + \sum_i \sum^3_{m=1} J_{m\perp} \left( \mathbf{S}_{i,1} \cdot \mathbf{S}_{i+\delta \mathbf{r}_m,1} + \mathbf{S}_{i,2} \cdot \mathbf{S}_{i+\delta \mathbf{r}_m,2}\right)  + \sum_i \sum^3_{m=1} J_{m\|} \mathbf{S}_{i,2} \cdot \mathbf{S}_{i+\delta \mathbf{r}_m,1} - h \sum_i \left(S^z_{i,1}+S^z_{i,2} \right),
  \label{ham1}
\end{equation}
where ${\bf S}_{i,k}$ denote $k$-th spin in $i$-th dimer ($k=1,2$), $z$ is a quantized axis along which the magnetic field $h$ is directed, and $\delta \mathbf{r}_m$ are vectors connected neighboring spins (see Fig.~\ref{Ba3Cr2O8}(a)).

We derive a Bose-analog of spin Hamiltonian \eqref{ham1} in the standard way \cite{sach,Dodds,kot} by introducing three Bose-operators $a$, $b$, and $c$ for each dimer which act on the vacuum spin state $|0\rangle=\frac{1}{\sqrt{2}}\left(|\uparrow\downarrow\rangle-|\downarrow\uparrow\rangle \right)$ as follows: $ a|0\rangle=b|0\rangle=c|0\rangle=0$, $a^+|0\rangle=|\uparrow\uparrow\rangle$, $b^+|0\rangle=|\downarrow\downarrow\rangle$, and $c^+|0\rangle=\frac{1}{\sqrt{2}}\left(|\uparrow\downarrow\rangle+|\downarrow\uparrow\rangle \right)$. One has for spin operators
\begin{equation}
  \begin{split}
    S^+_{i,1}=\frac{1}{\sqrt{2}}(a_i^+(c_i-1)+(c_i^++1)b_i),\\
    S^+_{i,2}=\frac{1}{\sqrt{2}}(a_i^+(c_i+1)+(c_i^+-1)b_i),\\
    S^z_{i,1}=\frac{1}{2}((c_i^++c_i)+a_i^+a_i-b_i^+b_i),\\
    S^z_{i,2}=\frac{1}{2}(-(c_i^++c_i)+a_i^+a_i-b_i^+b_i).
  \end{split}
  \label{spinrep}
\end{equation}
To fulfill the requirement that no more than one triplon $a$, $b$ or $c$ can sit on the same bond, one has to introduce constraint terms into the Hamiltonian which describe an infinite repulsion between triplons $U \sum_i(a^+_i a^+_i a_i a_i + b^+_i b^+_i b_i b_i + c^+_i c^+_i c_i c_i + a^+_i b^+_i a_i b_i + a^+_i c^+_i a_i c_i + b^+_i c^+_i b_i c_i)$, where $U \to +\infty$.

Substituting Eqs.~\eqref{spinrep} into Eq.~\eqref{ham1} and taking the Fourier transform we obtain for the Hamiltonian up to a constant
\begin{eqnarray}
  \mathcal{H} &=& \mathcal{H}_2+\mathcal{H}_3+\mathcal{H}_4,\\
\label{h2}
	\mathcal{H}_2 &=& \sum_\mathbf{p} \left[ \left( J - h + J_2(\mathbf{p})\right)a^+_\mathbf{p} a_\mathbf{p} + \left( J + h + J_2(\mathbf{p})\right)b^+_\mathbf{p} b_\mathbf{p}- J_2(\mathbf{p}) (a_\mathbf{p} b_{-\mathbf{p}} + a^+_\mathbf{p} b^+_{-\mathbf{p}})  \right.\nonumber\\
	&&{}\left.  + \left( J + J_2(\mathbf{p})\right)c^+_\mathbf{p} c_\mathbf{p} + \frac{J_2(\mathbf{p})}{2}(c_\mathbf{p} c_{-\mathbf{p}}+c^+_\mathbf{p} c^+_{-\mathbf{p}})\right],\\
\label{h3}
	 \mathcal{H}_{3} &=& \frac{i}{2\sqrt{N}} \sum_{\mathbf{p}_1=\mathbf{p}_2+\mathbf{p}_3} \left( a^+_1a_2c_3(J_3(\mathbf{p}_2) - J_3(\mathbf{p}_3)) + b^+_1b_2c_3(J_3(\mathbf{p}_3) - J_3(\mathbf{p}_2)) + c^+_1a_2b_3(J_3(\mathbf{p}_2) - J_3(\mathbf{p}_3)) \right)  \nonumber\\
	&&{} + \frac{i}{2\sqrt{N}} \sum_{\mathbf{p}_1+\mathbf{p}_2=\mathbf{p}_3} \left( a^+_1c^+_2a_3(J_3(\mathbf{p}_2) - J_3(\mathbf{p}_1)) + b^+_1c^+_2b_3(J_3(\mathbf{p}_1)- J_3(\mathbf{p}_2)) + a^+_1b^+_2c_3(J_3(\mathbf{p}_2 ) - J_3(\mathbf{p}_1)) \right),\\
\label{h4}
\mathcal{H}_4 &=& \frac{1}{N}\sum_{\mathbf{p_1}+\mathbf{p_2}=\mathbf{p_3}+\mathbf{p_4}}
    \left[\left( U +  \frac12J_4({\mathbf{p}_1-\mathbf{p}_3}) \right)(a^+_1a^+_2a_3a_4+
    b^+_1b^+_2b_3b_4) + U c^+_1c^+_2c_3c_4
		+ \left( U - J_4({\mathbf{p}_1-\mathbf{p}_3}) \right)
    a^+_1b^+_2a_3b_4  \right. \nonumber\\
		&&{}\left.+ \left( U + J_4({\mathbf{p}_1-\mathbf{p}_4}) \right)(a^+_1c^+_2a_3c_4+
    b^+_1c^+_2b_3c_4) + J_4({\mathbf{p}_1-\mathbf{p}_4})(c^+_1c^+_2a_3b_4+
    a^+_1b^+_2c_3c_4)
    \right],
\end{eqnarray}
where $N$ is the number of dimers in the system, we omit some indexes $\bf p$ in Eqs.~\eqref{h3} and \eqref{h4},
\begin{eqnarray}
  J_2(\mathbf{p}) &=& J_{1\perp} \cos{p_a}+J_{2\perp} \cos{p_b}+J_{3\perp} \cos{(p_a+p_b)}-\frac{J_{1\|}}{2}\cos{p_c}-\frac{J_{2\|}}{2}\cos{(p_c-p_a)}-\frac{J_{3\|}}{2}\cos{(p_c-p_a-p_b)},\\
	\label{j3}
	J_3(\mathbf{p}) &=& J_{1\|}\sin{p_c}+J_{2\|}\sin{(p_c-p_a)}+J_{3\|}\sin{(p_c-p_a-p_b)},\\
	J_4(\mathbf{p}) &=& J_{1\perp} \cos{p_a}+J_{2\perp} \cos{p_b}+J_{3\perp} \cos{(p_a+p_b)}+\frac{J_{1\|}}{2}\cos{p_c}+\frac{J_{2\|}}{2}\cos{(p_c-p_a)}+\frac{J_{3\|}}{2}\cos{(p_c-p_a-p_b)},
\end{eqnarray}
and $p_{a,b,c}$ are projections of $\bf p$ on the corresponding axes (see Fig.~\ref{Ba3Cr2O8}).

Taking into account that $[ {\cal H}, \sum_i(S_{i,1}^z +S_{i,2}^z) ] =0$ and using Eqs.~\eqref{spinrep} one concludes that $h$ and $-h$ play the role of chemical potentials for $a$ and $b$ triplons. Then, one can easily find spectra of triplons at $h\le h_c$ using those at $h=0$: the spectrum of triplons $c$ is not effected by magnetic field and spectra of $a$ and $b$ triplons are shifted by $h$ downwards and upwards, respectively.

\subsection{Triplons spectra at zero field}
\label{spectra}

To find spectra of triplons $a$, $b$ and $c$ as a series in the interdimer coupling, it is convenient to introduce the following Green's functions (cf.\ Ref.~\cite{Sizanov}):
\begin{eqnarray}
\label{g}
G_a(p) &=&-i\langle a_p a^+_p\rangle, \quad G_b(p)=-i\langle b_p b^+_p\rangle,\\
\label{fab}
  F_{ab}(p) &=& -i\langle b^+_{-p} a^+_p\rangle, \quad   \overline{F}_{ab}(p) = -i\langle a_{-p} b_p\rangle,\\
\label{fc}
G_c(p) &=& -i\langle c_p c^+_p\rangle, \quad  F_c(p) = -i\langle c^+_{-p} c^+_p\rangle, \quad   \overline{F}_c(p) = -i\langle c_{-p} c_p\rangle,
\end{eqnarray}
where $p=(\omega, \mathbf{p})$. Notice that $G_a(p)=G_b(p)=G_c(p)$ at zero field. A set of Dyson equations for $G_a(p)$ and $F_{ab}(p)$ has the form
\begin{equation}
  \begin{split}
    &G_a(p)=G_{0a}(p)+G_{0a}(p) \Sigma_p G_a(p) + G_{0a}(p) \Pi_p F_{ab}(p), \\
    &F_{ab}(p)=G_{0a}(-p)\overline{\Pi}_pG_a(p)+G_{0a}(-p)\Sigma_{-p}F_{ab}(p),
  \end{split}
  \label{Dyson}
\end{equation}
where $G_{0a}(p)=(\omega-\varepsilon_{1\mathbf{p}}+i0)^{-1}$,
\begin{equation}
  \varepsilon_{1\mathbf{p}}=J+J_2({\mathbf{p}}),
  \label{spec1}
\end{equation}
$\Sigma_p$ is a normal self-energy part, $\Pi_p$ and $\overline{\Pi}_p$ are anomalous self-energy parts. Similar sets of equations can be constructed for couples $G_c(p)$, $F_c(p)$ and $\overline{G}_b(-p)=-i\langle b^+_{p} b_{p}\rangle$, $\overline{F}_{ab}(p)$. The solution of Eqs.~(\ref{Dyson}) has the form
\begin{eqnarray}
  \label{gf}
  G_a(p) &=& \frac{\omega+\varepsilon_{1\mathbf{p}}+\Sigma_{-p}}{D(p)}, \quad F_{ab}(p) =-\frac{\overline{\Pi}_p}{D(p)},\\
  \label{d}
	D(p) &=& \omega^2-\varepsilon^{2}_{1 \mathbf{p}}-\varepsilon_{1\mathbf{p}}\left( \Sigma_p + \Sigma_{-p} \right) + \omega \left( \Sigma_{-p} - \Sigma_{p} \right) - \Sigma_p \Sigma_{-p}+|\Pi_p|^2.
\end{eqnarray}
It is shown below that self-energy parts give corrections of at least second order in the interdimer interaction. Then, Eq.~\eqref{spec1} gives the first order spectrum of triplons $a$. The first order spectra for $b$ and $c$ triplons, which can be found from solutions of the corresponding sets of Dyson equations, are also given by Eq.~\eqref{spec1}.

\begin{figure}
  \noindent
  \hfil
  \includegraphics[scale=0.6]{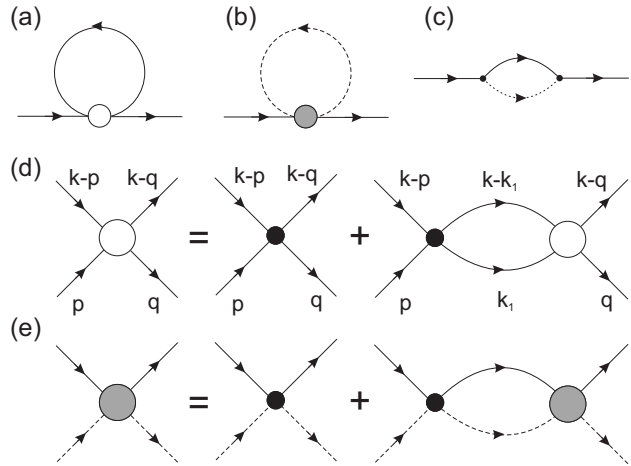}
  \hfil
  \caption{
(a)--(c) Diagrams giving the second order contributions to the normal self-energy part. Solid and dashed lines distinguish Green's functions of different triplons. (d) and (e) Equations for vertexes which appear in diagrams (a) and (b). Black dots stand for bare vertexes given by Eqs.~\eqref{h3} and \eqref{h4}.
}
  \label{diag}
\end{figure}

Second order corrections to spectra are defined by Hartree-Fock-type diagrams with zero-order vertexes and loop-type diagrams with bare Green's functions which are presented in Fig.~\ref{diag}. To find the zero-order vertexes $\Gamma^{(0)}$ one has to solve Bethe-Salpeter equations which are shown in Fig.\ref{diag}(d) and (e) and which give in the leading order \cite{Sizanov} (see Appendix~\ref{append} for more details)
\begin{equation}
\label{ver}
  \begin{split}
    \Gamma^{(0)}_a(p,k,q)=\Gamma^{(0)}_b(p,k,q)=\Gamma^{(0)}_c(p,k,q)=J-\omega_k, \\
    \Gamma^{(0)}_{ab}(p,k,q)=\Gamma^{(0)}_{bc}(p,k,q)=\Gamma^{(0)}_{ac}(p,k,q)=2(J-\omega_k),
  \end{split}
\end{equation}
where indexes denote the type of incoming (and outgoing) triplons. One obtains using Eqs.~\eqref{ver} for the contribution to the normal self-energy part from Hartree-Fock diagrams (see Fig.~\ref{diag}(a), (b))
\begin{equation}
\label{hf}
  \Sigma^{(2)HF}_p
	=\frac{2}{N J} \sum_{\mathbf{q}} |\Pi_q|^2
	= \frac{4\left(J_{1\perp}^2 + J_{2\perp}^2 + J_{3\perp}^2\right)+J_{1\|}^2 + J_{2\|}^2 +J_{3\|}^2}{4 J},
\end{equation}
where we use the first order expression for $\Pi_q=J_2({\bf q})$ that is given by Eq.~\eqref{h2}.
The second order contribution to the normal self-energy part from the loop diagram (see Fig.~\ref{diag}(c)) has the form
\begin{equation}
\label{loop}
  \begin{split}
    \Sigma^{(2)L}_p&=-\frac{1}{4 N J} \sum_\mathbf{q} (J_3(\mathbf{q})-J_3(\mathbf{p}-\mathbf{q}))^2\\ &=-\frac{J^2_{1\perp}(1+\cos{p_c})+J^2_{2\perp}(1+\cos{(p_c-p_a)})+J^2_{3\perp}(1+\cos{(p_c-p_a-p_b)})}{4 J}.
  \end{split}
\end{equation}
As a result we have from Eqs.~\eqref{d}, \eqref{hf}, and \eqref{loop} for triplons spectra in the second order
\begin{equation}
  \varepsilon_{2\mathbf{p}}= J + J_2({\mathbf{p}}) +  \Sigma^{(2)HF}_p + \Sigma^{(2)L}_p -\frac{J_2({\mathbf{p}})^2}{2 J}.
  \label{spec2}
\end{equation}

\subsection{Effective interaction between triplons $a$ at $h=h_c$}
\label{effint}

To describe the quantum phase transition to the magnetically ordered phase one has to find the effective interaction between triplons $a$ at $h=h_c$ which condense at this transition. The equation which gives the vertex $\Gamma_a(p)=\Gamma_a(p,0,k_0)$ in the first order in the interdimer interaction is presented in Fig.~\ref{diag}(d), where $k_0=(0,{\bf k}_0)$ and ${\bf k}_0$ is the momentum at which $\varepsilon_{2{\bf k}}$ reaches its minimum. The equation has the following explicit form:
\begin{equation}
\label{eqa}
    \Gamma_a^{(1)}(p)=\left( U + \frac{1}{2}J_4({\mathbf{p}+\mathbf{k}_0}) \right) - \frac{1}{N} \sum_{\mathbf{q}} \frac{\Gamma_a^{(1)}(q)}{\varepsilon_{1\mathbf{q}}-\Delta}\left(U + \frac{J_4({\mathbf{p}+\mathbf{q}}) + J_4({\mathbf{p}-\mathbf{q}})}{4}\right),
\end{equation}
where $\Delta=\varepsilon_{2{\bf k}_0}=h_c$ is the value of the spectrum gap at $h=0$. The solution of Eq.~\eqref{eqa} is quite cumbersome and we do not present it here.

\begin{figure}
  \noindent
  \hfil
  \includegraphics[scale=0.6]{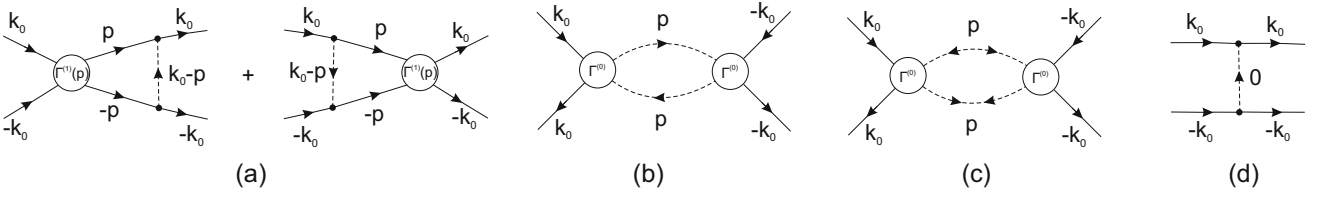}
  \hfil
  \caption{
  Diagrams contributing to the vertex $\Gamma_a(p=k_0)$ in the second order in the interdimer interaction. (a) The dashed and solid lines correspond to the bare $G_c(p)$ and $G_a(p)$, respectively. (b) Dashed lines stand for $G_b(p)$ or $G_c(p)$ of the second order, given by Eqs.~\eqref{gb2} and \eqref{gc2}, respectively. (c) Dashed lines correspond to anomalous functions $F_c(p)$ and $\overline{F}_c(p)$ given by Eqs.~\eqref{fc}. $\Gamma_a^{(1)}(p)$ is the solution of Eq.~\eqref{eqa}. $\Gamma^{(0)}$ represents zero-order vertexes \eqref{ver}.}
  \label{vertex2}
\end{figure}

Diagrams contributing to the vertex in the second order are shown in Fig.~\ref{vertex2}. One has to use the following second order normal Green's functions to calculate diagrams \ref{vertex2}(b):
\begin{eqnarray}
\label{gb2}
  G_b(p) &=& \frac{\omega+\varepsilon_{1\bf p}-h + \Sigma^{(2)}_{-p}}{(\omega-\varepsilon_{2\bf p}-h+i0)(\omega+\varepsilon_{2\bf p}-h-i0)},\\
	\label{gc2}
	G_c(p) &=& \frac{\omega+\varepsilon_{1\bf p}+\Sigma^{(2)}_{-p}}{(\omega-\varepsilon_{2\bf p}+i0)(\omega+\varepsilon_{2\bf p}-i0)}.
\end{eqnarray}
Vertexes $\Gamma^{(0)}$ in Fig.~\ref{vertex2}(b),(c) are given by Eqs.~\eqref{ver}. As a result of simple but tedious calculations one has for the second order correction to the vertex
\begin{equation}
\label{deqa}
  \delta \Gamma^{(2)}_a(p=k_0)=\frac{1}{4 J N} \sum_{\mathbf{q}} \frac{(J_3(\mathbf{k}_0-\mathbf{q})-J_3(\mathbf{q}))J_3(\mathbf{k}_0-\mathbf{q})}{\varepsilon_{1\bf q}-h}\Gamma_a^{(1)}(\mathbf{q})
	+\frac{3}{2JN} \sum_\mathbf{q} J_2(\mathbf{q})^2,
\end{equation}
where the first and the second terms stem from diagrams shown in Fig.~\ref{vertex2}(a) and (b)--(c), respectively. For simplicity, we assume here that components of ${\bf k}_0$ are equal either to $\pi$ or to 0 (as it is the case in $\rm Ba_3 Cr_2 O_8$ with ${\bf k}_0=(\pi,\pi,\pi)$). It is seen from Eqs.~\eqref{h3} and \eqref{j3} that the diagram presented in Fig.~\ref{vertex2}(d) gives zero at such ${\bf k}_0$.

\subsection{Triplons condensation}
\label{condensation}

Triplons $a$ condense at $h>h_c$ because their spectrum becomes negative. Then, one has to make the following shift at $h>h_c$: \cite{popov}
\begin{equation}
  a_{\mathbf{k}_0}\mapsto\sqrt{\rho N} e^{i \phi}+a_{\mathbf{k}_0},
   \label{cond1}
\end{equation}
where $\rho$ is the condensate density and $\phi$ is an arbitrary phase. It is seen from Eq.~\eqref{h2}, however, that terms linear in $b_{{\bf k}_0}$ and $b^+_{{\bf k}_0}$ appear in the Hamiltonian after shifting \eqref{cond1}. As a result one has to perform an extra shift to cancel these linear terms
\begin{equation}
    b_{\mathbf{k}_0}\mapsto\sqrt{\sigma N} e^{i \psi}+b_{\mathbf{k}_0}.
  \label{cond2}
\end{equation}
It is shown below that $\sigma$ is smaller than $\rho$ by a factor of $J_2({\bf k}_0)^2/J^2$ (see Eq.~\eqref{sigma}).

One has to minimize the free energy to find $\rho$, $\sigma$, $\phi$ and $\psi$ that has the form  in the second order at $T=0$
\begin{equation}
  \begin{split}
    \frac{E}{N}=&\left(\Delta+\frac{J_2({\mathbf{k}_0})^2}{2 J}-h\right) \rho + 2J\sigma -2 J_2({\bf k}_0) \cos{(\psi+\phi)} \sqrt{\rho \sigma}
		+\frac{\Gamma}{2}\rho^2 +2 J \rho \sigma,
  \end{split}
  \label{e0}
\end{equation}
where $\Gamma=2 \Gamma_a^{(2)}(k_0)$. Minimizing $E$ with respect to $\sigma$, $\phi$ and $\psi$ we obtain
\begin{equation}
  \begin{split}
	J_2({\bf k}_0)\cos(\psi + \phi)=|J_2({\bf k}_0)|,\\
  \sqrt{\sigma}=\frac{|J_2({\bf k}_0)|}{2J}\sqrt{\rho}.
	  \end{split}
		  \label{sigma}
\end{equation}
Substituting Eqs.~\eqref{sigma} into Eq.~\eqref{e0} one has in the second order in $J_2({\bf k}_0)/J$
\begin{equation}
  \begin{split}
    \frac{E}{N}=&\left(\Delta-h\right) \rho
		+\frac{1}{2}\left(\Gamma+\frac{J_2({\bf k}_0)^2}{J}\right)\rho^2
  \end{split}
  \label{e}
\end{equation}
that gives $\rho=(h-\Delta)(\Gamma + J_2({\bf k}_0)^2/J)^{-1}$.

\begin{figure}
  \noindent
  \hfil
  \includegraphics[scale=0.6]{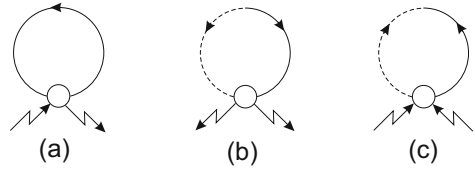}
  \hfil
  \caption{Diagrams giving temperature corrections to the free energy at $h>h_c$. Zigzag lines correspond to condensed triplons. Solid line and those with two arrows stand for $G_a(p)$ and $F_{ab}(p)$, $\overline{F}_{ab}(p)$ given by Eqs.~\eqref{g} and \eqref{fab}, respectively.}
  \label{T-corr}
\end{figure}

There are temperature corrections to $E$ coming from diagrams depicted in Fig.~\ref{T-corr} which have the form
\begin{eqnarray}
\label{dte}
   \frac{\delta_T E}{N}&=&2n(\Gamma \rho + J \sigma) + 2n \sqrt{\rho\sigma}\cos(\psi + \phi) J_2({\bf k}_0), \\
	n&=&\frac{1}{(2 \pi)^3}\int d^3 {\bf k} \left(e^{\varepsilon_{2\bf k}/T}-1\right)^{-1},
\end{eqnarray}
where the first and the second terms in Eq.~\eqref{dte} stem from diagrams shown in Fig.~\ref{T-corr}(a) and (b)--(c), respectively, and $n\propto T^{3/2}$ at $h=h_c(T)$ and small enough $T$. \cite{popov} Minimizing the free energy using Eqs.~\eqref{e} and \eqref{dte} we lead to the following expression:
\begin{equation}
\label{rho}
  \rho=\frac{h-\Delta-2\left( \Gamma + \frac{3 J_2({\bf k}_0)^2}{4J} \right)n }{\Gamma+J_2({\bf k}_0)^2/J}.
\end{equation}
One finds for temperature dependence of the critical field from Eq.~\eqref{rho}
\begin{equation}
\label{hct}
  h_c(T)=\Delta + 2\left( \Gamma + \frac{3 J_2({\bf k}_0)^2}{4J} \right)n.
\end{equation}

Another important characteristic of the phase transition is the staggered magnetization per spin $M_\perp$. One obtains for its square in the second order in $J_2({\bf k}_0)/J$ using Eqs.~\eqref{spinrep}, \eqref{cond1}, \eqref{cond2} and  \eqref{sigma}
\begin{equation}
\label{mperp}
  M_\perp^2=\frac{\rho-2\sqrt{\rho \sigma} \cos{(\psi + \phi)}+\sigma}{2}
	= \frac{\rho}{2} \left( 1 - \frac{J_2({\bf k}_0)}{2J}\right)^2.
\end{equation}

\section{Application to $\rm Ba_3 Cr_2 O_8$}
\label{appl}

\begin{figure}
  \noindent
  \hfil
  \includegraphics[scale=0.3]{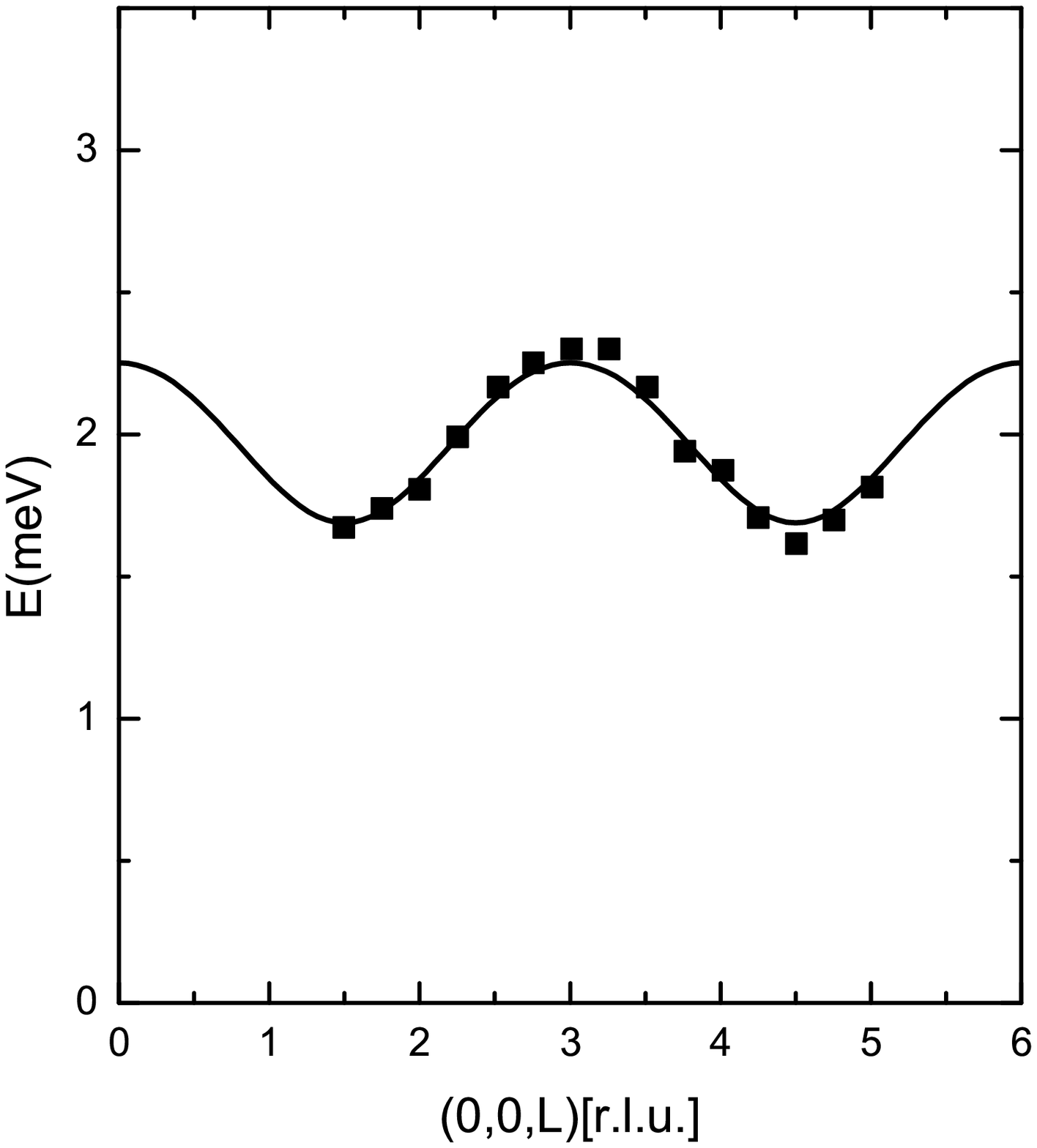}
  \includegraphics[scale=0.3]{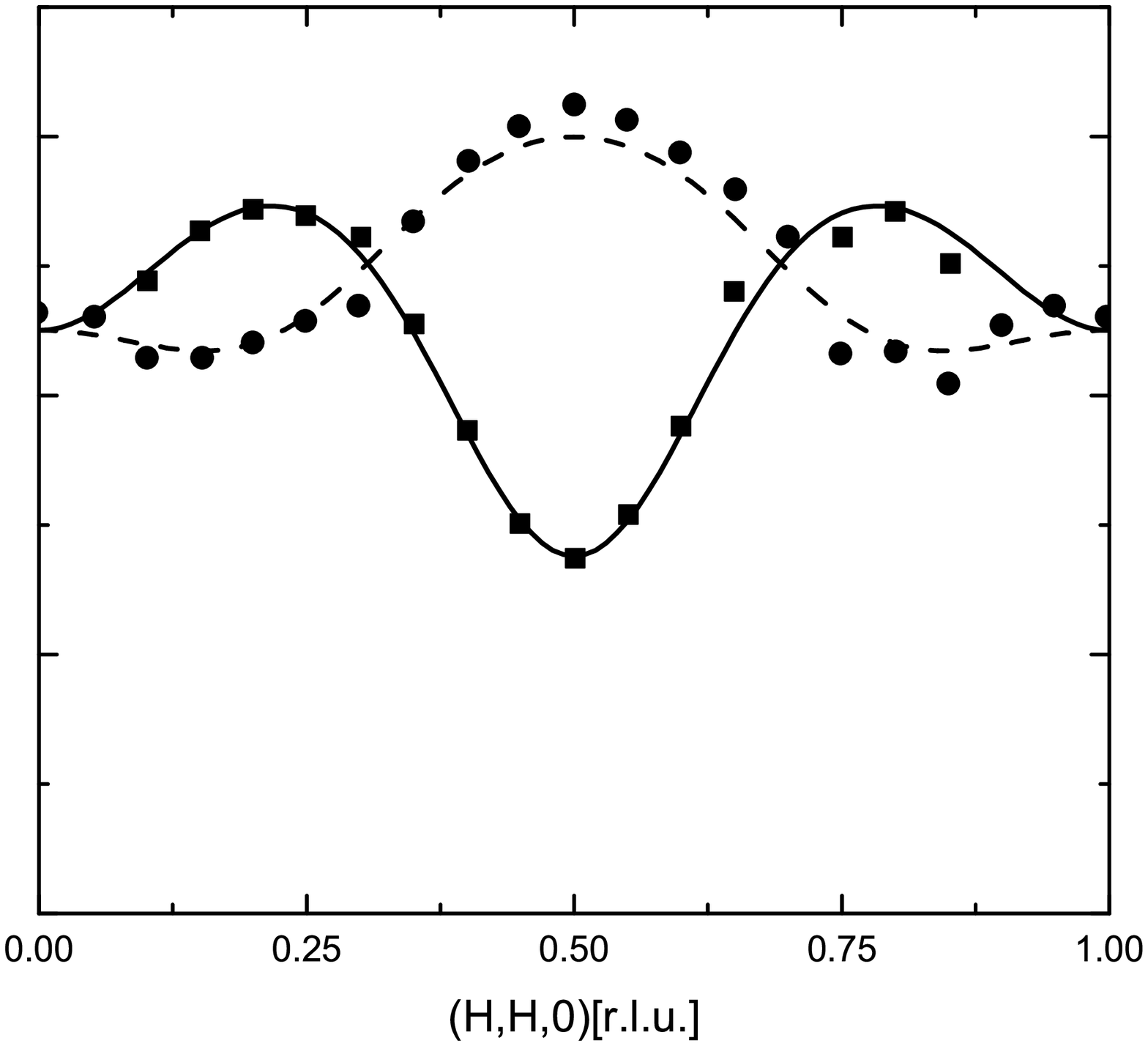}
	\includegraphics[scale=0.3]{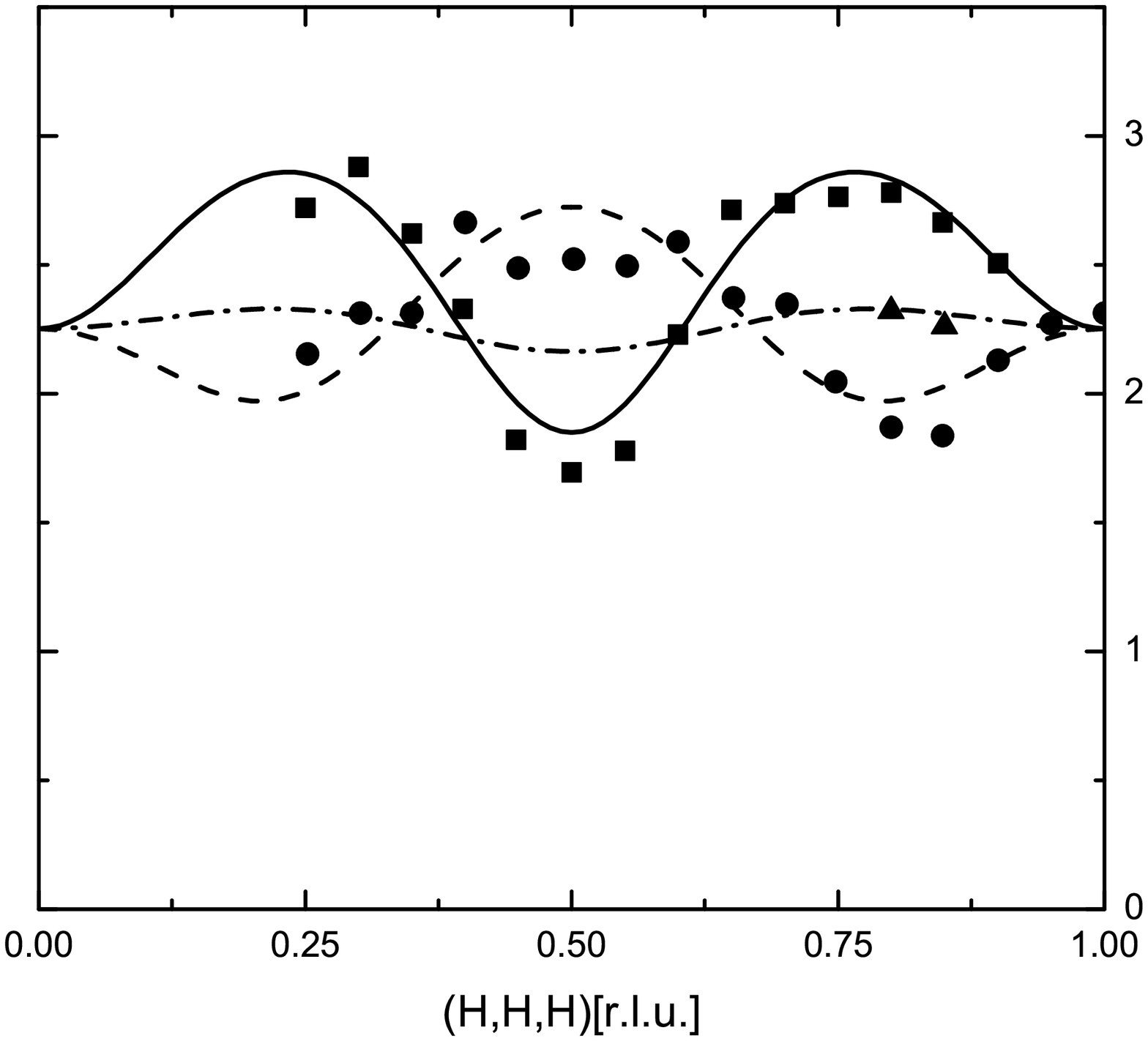}
  \hfil
  \caption{Triplon spectra in $\rm Ba_3 Cr_2 O_8$ along three directions in the Brillouin zone at zero field. Solid curves are drawn using Eqs.~\eqref{spec2} and \eqref{ec}. Dots are experimental data taken from Ref.~\cite{Kofu}. Two and three curves on some plots correspond to contributions from three crystal domains (see the text).}
  \label{specfig}
\end{figure}

The result is presented in Fig.~\ref{specfig} of our triplons spectrum best fit of neutron data\cite{Kofu} obtained in $\rm Ba_3 Cr_2 O_8$ at $h=0$. We have used Eq.~\eqref{spec2} for the spectrum. The corresponding exchange coupling constants expressed in meV are the following:
\begin{equation}
  \begin{split}
J=&2.18,\\
J_{1\perp}=&0.21, \quad J_{2\perp}=-0.06, \quad J_{3\perp}=-0.43,\\
J_{1\|}=&0.04, \quad J_{2\|}=-0.07, \quad J_{3\|}=-0.53.
  \end{split}
  \label{ec}
\end{equation}
The spectrum has a minimum at the momentum $\mathbf{k}_0=(\pi,\pi,\pi)$ that is in agreement with the experimental data. \cite{KofuExp} Two and three curves on some plots in Fig.~\ref{specfig} correspond to contributions from three crystal domains observed experimentally~\cite{Kofu} in which exchange constants are permuted as follows: $\{J_{1\perp,\|},J_{2\perp,\|},J_{3\perp,\|}\}\to\{J_{2\perp,\|},J_{3\perp,\|},J_{1\perp,\|}\}\to\{J_{3\perp,\|},J_{1\perp,\|},J_{2\perp,\|}\}$ (see Ref.~\cite{Kofu} for details).

Taking into account the experimentally obtained $g$-factor \cite{zvyag,KofuExp} $g=1.94$ we have for critical fields values
\begin{equation}
  \begin{split}
h_c &= \varepsilon_{2{\bf k}_0}\approx12.5~{\rm T},\\
h_s &= 2S(J+2J_{1\perp}+J_{1\|})\approx23.5~{\rm T},
  \end{split}
  \label{hcs}
\end{equation}
which are in excellent agreement with corresponding values of 12.5~T and 23.6~T obtained experimentally at $T=0.3$~K and ${\bf h}\|\hat c$. \cite{Aczel} It should be noted that critical fields values depend on the field direction and differ by about $0.3\div0.6$~T (they vary also in different experiments in the same range). This diversity is attributed to a small anisotropy of the order of 0.03~meV that we do not consider here.

\begin{figure}
  \noindent
  \hfil
  \includegraphics[scale=0.3]{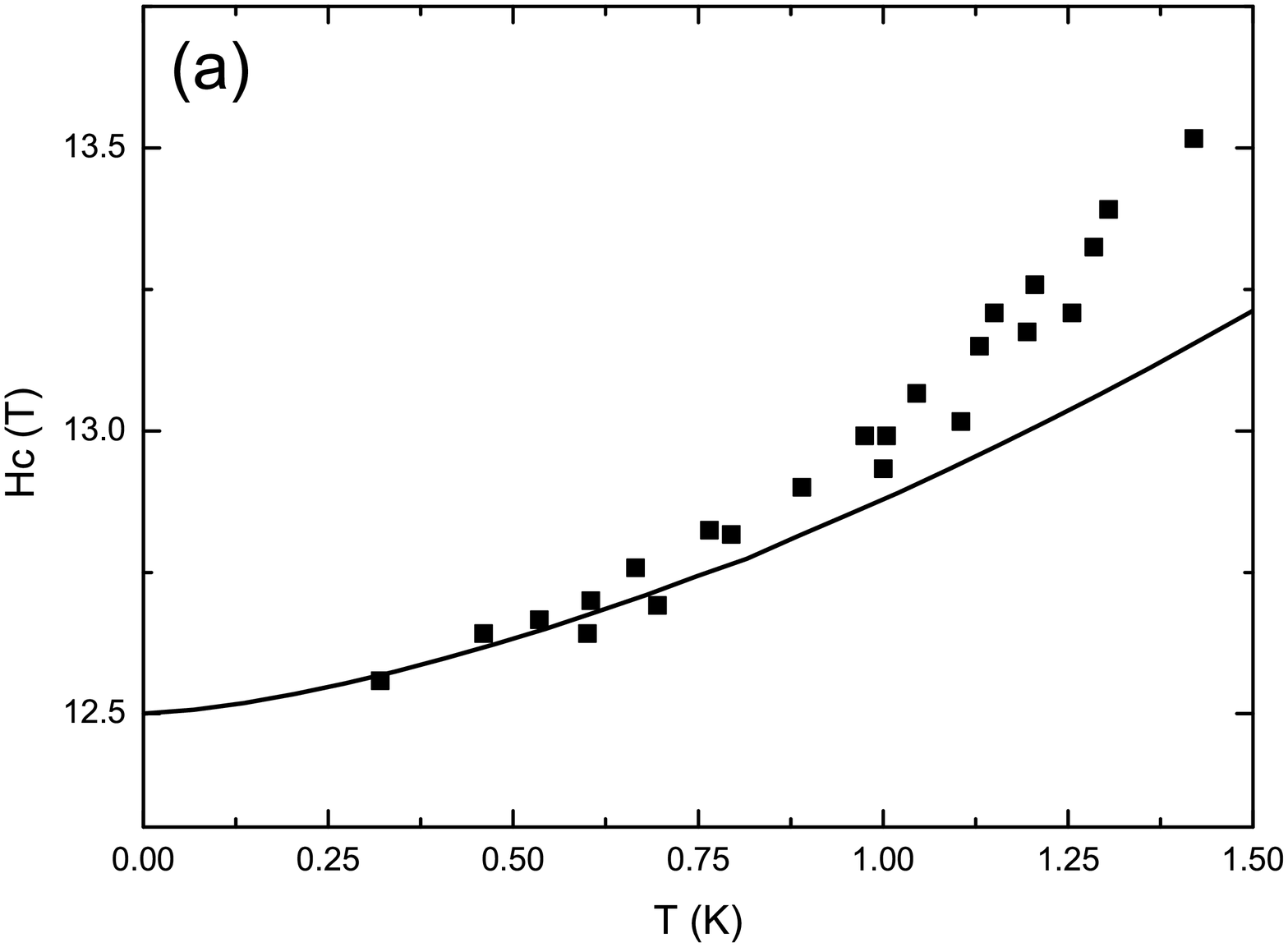}
	  \includegraphics[scale=0.3]{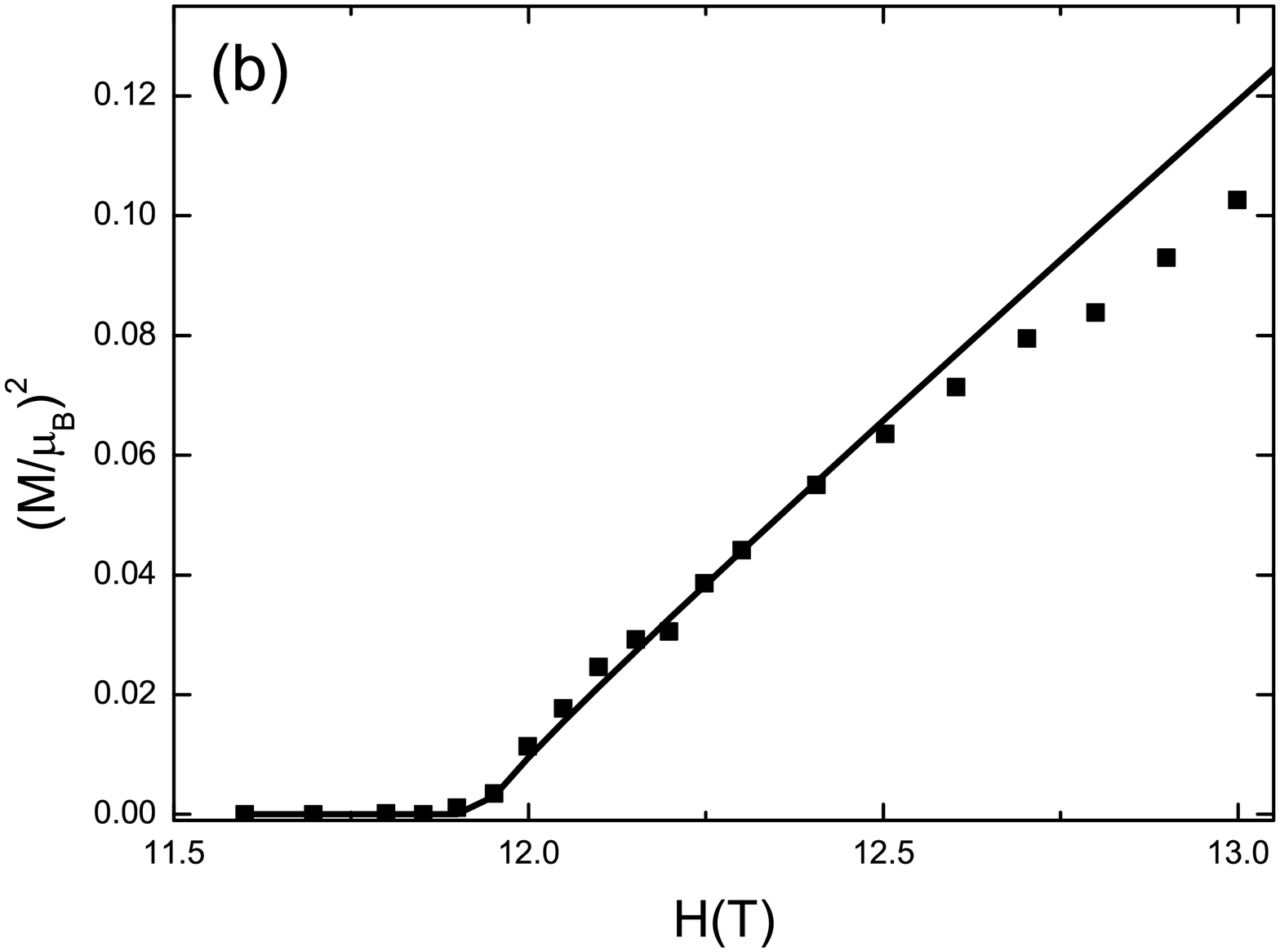}
  \hfil
  \caption{(a) Temperature dependence of the critical field in $\rm Ba_3 Cr_2 O_8$. The solid line is drawn using Eq.~\eqref{hct} and points are experimental data from Ref.~\cite{Aczel}. (b) Comparison of theoretical result \eqref{mperp} with experimental data \cite{KofuExp} for the square of staggered magnetization at $h>h_c$ and $T=0.2$~K.}
  \label{hcmag}
\end{figure}

The solution of Eq.~\eqref{eqa} and Eq.~\eqref{deqa} give for the effective interaction between triplons $a$ in the first and the second order $\Gamma_a^{(1)}(k_0)\approx0.27$~meV and $\Gamma_a^{(2)}(k_0)\approx0.17$~meV, respectively. Results are presented in Fig.~\ref{hcmag}(a) of our calculation of $h_c(T)$ using Eq.~\eqref{hct}. A reasonable agreement with experiment is seen at $T<1$~K. Notice that the spectrum has the following form near its minimum:
\begin{equation}
	\epsilon_{2\bf k}\approx \Delta-h+A_1k_1^2+A_2k_2^2+A_3k_3^2,
\end{equation}
where $A_1\approx11.8$~K, $A_2\approx1.1$~K, $A_3\approx0.17$~K and indexes 1--3 enumerate axes in the suitable local basis in the $\bf k$-space. Thus, the small stiffness constant $A_3$ restricts the range of validity of the 3D BEC law $h_c(T)\propto T^{3/2}$ to quite a small interval of $T$ that is bounded by 1~K. This conclusion is in accordance with those of experimental papers \cite{Aczel,KofuExp}. Besides, one expects also that our theoretical discussion presented above becomes invalid in $\rm Ba_3 Cr_2 O_8$ at $T>1$~K due to large temperature fluctuations which are not taken into account. \cite{popov} The discrepancy is attributed to this fact between theory and experiment in Fig.~\ref{hcmag}(a) that is pronounced at $T\agt1$~K.

The staggered magnetization square as a function of $h$ is drawn in Fig.~\ref{hcmag}(b) that shows good agreement at $(h-h_c)\alt0.06h_c$ between Eq.~\eqref{mperp} and results of elastic neutron scattering \cite{KofuExp} at $T=0.2$~K.

\section{Summary}
\label{sum}

To conclude, we develop a theory describing triplon spectra and the quantum field-induced phase transition to a magnetically ordered state in dimer systems containing stacked triangular layers (see Fig.~\ref{Ba3Cr2O8}). Triplon spectra and effective interaction between triplons at $h=h_c$ are derived in the second order in the interdimer interaction. Expressions for the condensed triplons density, $h_c(T)$ and the staggered magnetization $M_\perp$ are derived. The proposed theory is applied to $\rm Ba_3 Cr_2 O_8$. Good agreement is achieved between the theory and experimentally obtained critical fields values, triplon spectra, $h_c(T)$ and $M_\perp$ if exchange constants are given by Eq.~\eqref{ec}. It is demonstrated that in accordance with experimental data the system follows the 3D BEC scenario at $T<1$~K only. This is a consequence of large anisotropy of the spectrum near its minimum.

\begin{acknowledgments}

This work is supported by RF President (Grant No.\ MD-274.2012.2), the Dynasty foundation and RFBR Grants No.\ 12-02-01234 and No.\ 12-02-00498.

\end{acknowledgments}

\appendix

\section{Zero-order vertexes}
\label{append}

We derive here Eqs.~\eqref{ver} for zero-order vertexes. The Bethe-Salpeter equation for $\Gamma^{(0)}_a(p,k,q)$ that is shown in Fig.\ref{diag}(d) is written as
\begin{equation}
\label{g0}
  \Gamma^{(0)}_a(p,k,q)=U+\frac{(-i)i^2 2 U }{N} \int \frac{d\omega_{k_1}}{2 \pi} \sum_{{\bf k}_1} \frac{\Gamma^{(0)}_a(k_1,k,q)}{(\omega_k-\omega_{k_1}-J+i0)(\omega_{k_1}-J+i0)}.
\end{equation}
Equations for $\Gamma^{(0)}_b$ and $\Gamma^{(0)}_c$ have the same form and those for $\Gamma^{(0)}_{ab}$, $\Gamma^{(0)}_{bc}$ and $\Gamma^{(0)}_{ac}$ differ from Eq.~\eqref{g0} by a factor of 2 in the second term in the right-hand side. It is seen from Eq.~\eqref{g0} that $\Gamma^{(0)}(p,k,q)$ depends only on $\omega_k$. As a result one obtains after simple integration over $\omega_{k_1}$
\begin{equation}
  \Gamma^{(0)}_a=U+\frac{U \Gamma^{(0)}_a}{\omega_k-J}, \quad
	\Gamma^{(0)}_{ab}=U+\frac{U \Gamma^{(0)}_{ab}}{2(\omega_k-J)}
\end{equation}
that gives Eqs.~\eqref{ver} at $U \to +\infty$.

\bibliography{bibliography}

\end{document}